# Simulating Organogenesis in COMSOL: Tissue Mechanics

M. D. Peters[*,1,2], D. Iber,[1,2]
[1]D-BSSE, ETH Zurich, Mattenstrasse 26, 4058 Basel, Switzerland, [2]Swiss Institute of Bioinformatics (SIB), Switzerland
[*]Corresponding author: michael.peters@bsse.ethz.ch

**Abstract:** During growth, tissue expands and deforms. Given its elastic properties, stresses emerge in an expanding and deforming tissue. Cell rearrangements can dissipate these stresses and numerous experiments confirm the viscoelastic properties of tissues [1]–[4]. On long time scales, as characteristic for many developmental processes, tissue is therefore typically represented as a liquid, viscous material and is then described by the Stokes equation [5]–[7]. On short time scales, however, tissues have mainly elastic properties. In discrete cell-based tissue models, the elastic tissue properties are realized by springs between cell vertices [8], [9]. In this article, we adopt a macroscale perspective of tissue and consider it as homogeneous material. Therefore, we may use the "Structural Mechanics" module in COMSOL Multiphysics in order to model the viscoelastic behavior of tissue. Concretely, we consider two examples: first, we aim at numerically reproducing published [10] analytical results for the sea urchin blastula. Afterwards, we numerically solve a continuum mechanics model for the compression and relaxation experiments presented in [4].

**Keywords:** Tissue mechanics, hyperelastic material, viscoelastic material.

## Introduction

In this article, we implement a continuum mechanics based approach to model deformations in biological tissues with the aid of COMSOL Multiphysics. In earlier works, we have already successfully employed COMSOL to simulate reaction diffusion models [11]–[16], particularly in view of parameter estimation [12], phase field based models for growing domains [13], cell based descriptions [14], image based geometry models [15] and deforming and interacting domains [16].

Continuum mechanical models are widely used to simulate the mechanical properties of organs, see e.g. [10], [17] and the references therein.
Tissue expands and deforms during growth. Thus, given its elastic properties, stresses emerge in an expanding and deforming tissue. Cell rearrangements can dissipate these stresses and numerous experiments confirm the viscoelastic properties of tissues [1]–[4]. On long time scales, as characteristic for many developmental processes, tissue is therefore typically represented as a liquid, viscous material and is then described by the Stokes equation [5]–[7]. On short time scales, however, tissues have mainly elastic properties. To represent these properties, essentially two different approaches exist. Besides the continuum mechanical description, there also exist discrete cell-based approaches. In discrete cell-based tissue models, each cell is represented by a massless elastic polygon, where the edges of a cell and the interconnections between cells are represented by elastic springs [6]. In addition, fluid can be represented as Newtonian fluid. The interaction between fluid and elastic structures is then accounted for by an immersed boundary method cf. [8], [9] and the references therein.

Within this article, we adopt a macroscale perspective of tissue and consider it consequentially as a homogeneous (visco-) elastic material. This permits a continuum mechanics description, which we can implement on COMSOL.

Concretely, we consider two examples: first, we aim at numerically reproducing analytical results from literature, see [10]. Afterwards, we consider a continuum mechanics based numerical model for the compression and relaxation experiments presented in [4].

## The sea urchin blastula

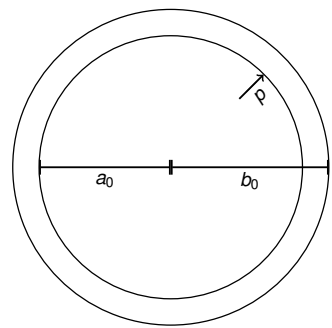

**Figure 1**. Model of the sea urchin blastula.

The blastula stage of the sea urchin development begins at the 128-cell stage. At this point, the cells form a hollow sphere surrounding a central cavity or *blastocoel*. Here, tight junctions and adhesion complexes connect the *blastomeres* into a seamless epithelium, which completely encloses the blastocoel, cf. [18]. The enclosed fluid exerts an outward pressure onto the epithelial cells. The cells mitigate the induced stress by moving and deforming, cf. [10]. It is assumed that this is one aspect which drives *gastrulation*, which is the phase in which the blastula reorganizes into a multilayered structure, cf. [19].

The blastula $B$ can be modelled by a hollow ball with radii $b_0 > a_0 > 0$ and thickness $h_0 = b_0 - a_0$. The interior boundary shall be denoted by $\Gamma_{int} := \{|x| = a_0\}$ and the exterior boundary by $\Gamma_{out} := \{|x| = b_0\}$. A visualization of the situation can be found in Figure 1. For simplicity, we neglect residual stresses and model the blastula as isotropic and incompressible. Thus, we consider the steady state equation
$$div\ \sigma = 0$$
with the pressure boundary condition $\sigma \cdot n = -pn$ at $\Gamma_{int}$ and a free boundary condition at $\Gamma_{out}$. Moreover, we consider a hyperelastic material law given by the constitutive equation
$$\sigma = J^{-1}\frac{\partial W(F)}{\partial F}F^{\mathsf{T}}$$
where the strain energy density function of Fung type is given by
$$W = \frac{C}{\alpha}[e^{\alpha(I_1-3)} - 1]$$
cf. [20], [21], with
$$C = \frac{E}{6} = 0.2 kPa, \qquad I_1 = trace(FF^{\mathsf{T}})$$
For $\alpha \to 0$, we have $W \to C(I_1 - 3)$, which constitutes a neo-Hookean material, cf. [10].

**Model Setup in COMSOL**

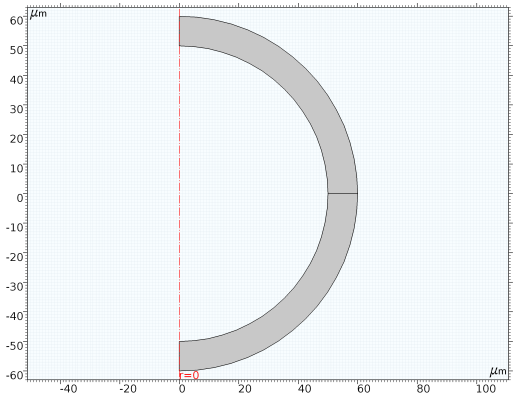

**Figure 2**. Computational domain for the blastula.

In order to reduce the computational cost, we consider only a single 2D slice of the rotational symmetric model, as shown in Figure 2. This is realized in COMSOL[1] by choosing a "2D Axisymmetric" component. The mesh for the finite element simulation is set to "Mapped".

In order to incorporate the physics, we employ the "Solid Mechanics" interface. The blastula is modelled by a "Hyperelastic Material" node, where we select a "User defined" material model and enable the "Nearly incompressible material" checkbox. Moreover, we set
$$W_{siso} = \frac{C}{\alpha}[e^{\alpha(\bar{I}_1-3)} - 1]$$
and
$$W_{svol} = 0.5\kappa(J-1)^2$$
where $\kappa = 2.2 GPa$ is the bulk modulus of water, $J = \det(F)$ the elastic volume ratio and $\bar{I}_1 = J^{-2/3}I_1$ the first invariant of the isochoric right Cauchy-Green deformation tensor. For the density, we use $\rho = 1000 kg/m^3$, i.e. the density of water. Note that we have $\bar{I}_1 \to I_1$ and $W_{svol} \to 0$ for the incompressible limit $J \to 1$.

An "Axial Symmetry" boundary condition is set automatically for the boundary points with $r = 0$. The outer boundary $\Gamma_{out}$ is not constrained, i.e. we use a "Free" boundary condition here. At the interior boundary $\Gamma_{int}$, we prescribe the pressure $p$ via a "Boundary Load" node. Finally, in order to suppress rigid body motions, we employ a "Fixed Displacement", i.e. we prescribe $z = 0$, for the line segment $(1-t)[a_0, 0]^{\mathsf{T}} + t[b_0, 0]^{\mathsf{T}}, 0 \leq t \leq 1$.

**Numerical results**

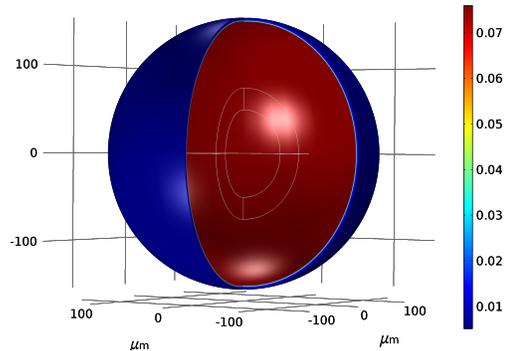

**Figure 3**. von Mises stress distribution (in $\mu N/\mu m^2$) for $h_0 = 25\mu m$, $\alpha = 0.2$, $p = 1kPa$.

---

[1]We use COMSOL Multiphysics 5.3 here.

We are mainly interested in reproducing the analytical computations from [10] numerically. Figure 3 shows the von Mises stress distribution inside the blastula wall for $a_0 = 50\mu m, b_0 = 75\mu m, \alpha = 0.2, p = 1kPa$. In order to obtain our results, we have used a "Stationary Study" with an "Auxiliary sweep" for the pressure $p$.

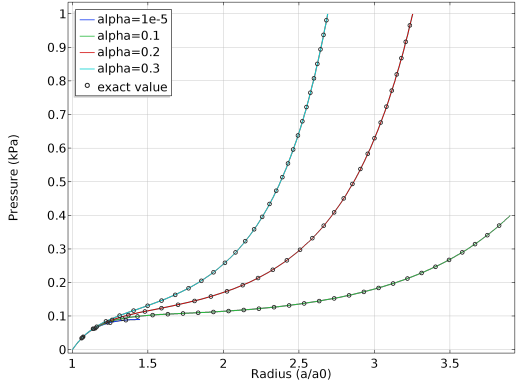

**Figure 4**. Parameter sweep for different values of $\alpha$. The colored curves indicate the numerically approximated pressure values, whereas the black circles indicate the exact pressure values.

Figure 4 shows the relative stretch of the blastula for $\alpha = 10^{-5}, 0.1, 0.2, 0.3$. The black circles indicate the exact values for the pressure computed via the analytical formula from [10]. As can be seen, the numerical results perfectly match the analytical solution.

Note that the material stiffens for increasing values of $\alpha$. As a consequence, a higher pressure is necessary in order to achieve the same stretch. We remark that the blastula would burst in the neo-Hookean limit case, i.e. $\alpha \to 0$, for pressures $p > 0.1kPa$, cf. [10]. This is in accordance with our simulation for $\alpha = 10^{-5}$, which crashes when the pressure approaches $0.1kPa$.

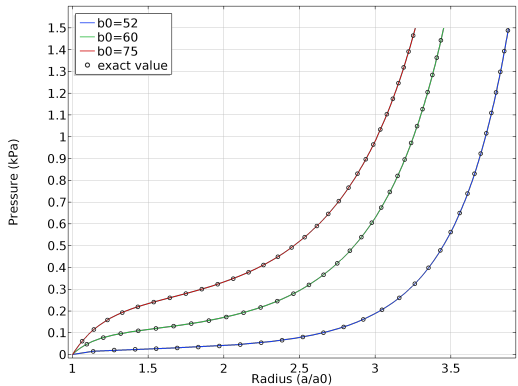

**Figure 5**. Parameter sweep for different values of $h_0$. The colored curves indicate the numerically approximated pressure values, whereas the black circles indicate the exact pressure values.

In Figure 5, we have depicted the stretch versus pressure curves for different thicknesses of the blastula, i.e. $a_0 = 50\mu m, b_0 = 52\mu m, 60\mu m, 75\mu m$. As in the previous figure, the black circles indicate the exact values for the pressure computed via the analytical solution from [10]. It turns out that the numerical results perfectly match the analytical solution here, as well.

Therefore, we may conclude that our implementation of the material law is feasible for biomechanical simulations.

**The multicellular spheroid**

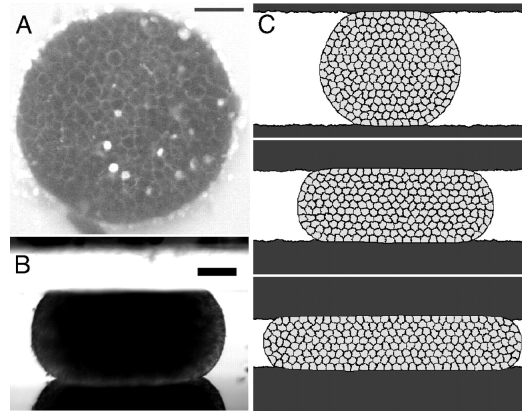

**Figure 6**. Shape of aggregates. (A) top view. (B) Side view. (C) Numerical simulation. Reprinted from "The role of fluctuations and stress on the effective viscosity of cell aggregates," by P. Marmottant *et al.*, *PNAS*, vol. 106, no. 41, pp. 17271–17275, 2001. Reprinted with the permission of *PNAS*.

Cell aggregates or *multicellular spheroids* are comprised either of a single cell type or of combinations of different cell types, see Figure 6A. They are used for *in vitro* studies of morphogenesis, cancer invasion and tissue engineering, cf. [4]. Particularly, it is possible to infer mechanical properties of a tissue directly from compression experiments. In [4] it is shown experimentally that tissue exhibits viscoelastic behavior on tissue level and might therefore be represented as homogenous viscoelastic material. This behavior is caused by topological rearrangements within the tissue in order to mitigate stresses, as it is observed e.g. in polymers as well, cf. [22].

Our goal here is to numerically reproduce the wet lab compression experiments conducted in [4] with the aid of COMSOL. In [4] a multicellular spheroid comprised of mouse embryonic carcinoma cells was compressed between two glass plates, see Figure 6B. We would like to emphasize that numerical experiments have been conducted in [4], as well. However, there a cellular pots model [23], [24] (cell based point of view) was employed rather than a continuum mechanics approach.

We represent the multicellular spheroid $M$ by a ball with radius $a_0 > 0$. In contrast to the previous example, we shall consider here the transient case
$$div\ \sigma = \rho \partial_{tt} u$$
where the mass density $\rho$ is constant over time due to the incompressibility of the material. Moreover, we take into account viscoelastic behavior by inserting the viscoelastic branches $\Psi_m(t)$, $m = 1, \dots, n$ into the strain energy function, i.e.
$$W_{ve} = W + \sum_{m=1}^{n} \Psi_m(t)$$
The functions $\Psi_m$ describe the *configurational free energy*, cf. [22].

**Model Setup in COMSOL**

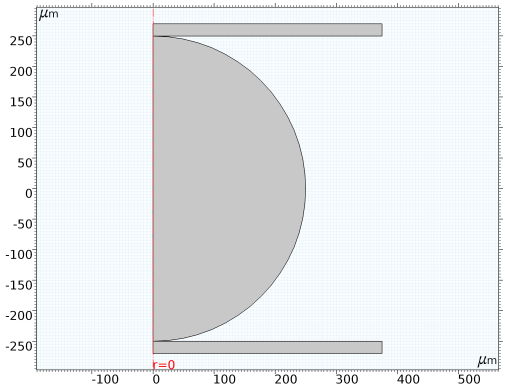

**Figure 7**. Computational domain for the multicellular spheroid.

As in the previous example, we exploit the rotational symmetry in our problem to reduce the computational cost and use a "2D Axisymmetric" component, see Figure 7. Besides the mechanical properties of the aggregate, we have to take into account the contact that arises between the two horizontal plates and the aggregate. This is achieved by the use of the "Form Assembly" node in the geometry definition. In addition, we define "Contact Pairs" between the upper plate and the spheroid as well as between the lower plate and the spheroid. In both cases, the spheroid's boundary is chosen as "Destination Boundary" and the plates' boundaries are used as "Source boundaries". For the finite element simulation, we use a "Mapped" mesh for the plates and a "Free Triangular" mesh for the spheroid. The mesh size at the spheroid's boundary is chosen as half the mesh size of the plates' contact boundaries.

In order to incorporate the physics, we use again the "Solid Mechanics" interface. The cell aggregate is modelled by a "Hyperelastic Material" with the same constitutive relations as in the previous example. Particularly, we use the same values for the density and the bulk modulus. In addition to the previous example, we incorporate now a "Viscoelasticity" node, where the "Material model"
is set to "Generalized Maxwell" with a single viscoelastic branch $\Psi_1$.

An "Axial Symmetry" boundary condition is set automatically for the boundary points with $r = 0$.
We fix the lower plate by employing a "Fixed Constraint"; for the upper plate, we describe the displacement via a "Prescribed Displacement" node.
For each of the contact pairs, we use a "Contact" boundary condition.

The "Contact Pressure Method" is set to "Penalty", the "Characteristic stiffness" is chosen as $E = 6C$, see previous example. The "Contact pressure penalty factor" is set to $40 \cdot solid.Eequ/solid.hmin\_dst$, which seems to be sufficient for moderate displacements of the upper plate. All remaining boundaries are set to "Free".

For the time dependent solver, we choose the "BDF Method" with a minimum order of 1 and a maximum order of 3. Finally, we set the radius of the ball to $a_0 = 500\mu m$.

**Numerical results**

First, we want to demonstrate the influence of the viscoelasticity concerning the behavior of the stress over time. To that end, we set the "Energy factor" for the "Viscoelasticity" node to $\beta_1 = 1.2$ and the "Relaxation time" to $\tau_1 = 200s$. The other parameters for the material law are chosen as in the previous example.

The upper image in Figure 8 shows the von Mises stress distribution inside the aggregate after compressing it within $0.2s$ to 80% of its original height. The graph in the upper part of Figure 9 shows how the average stress increases correspondingly.
The compression has been achieved by linearly displacing the upper plate in $-z$ direction. The lower image displays the stress distribution after $t = 1497s$. As can be seen, the stress has been released due to the viscoelasticity. The corresponding stress release curve is depicted in the lower part of Figure 9.

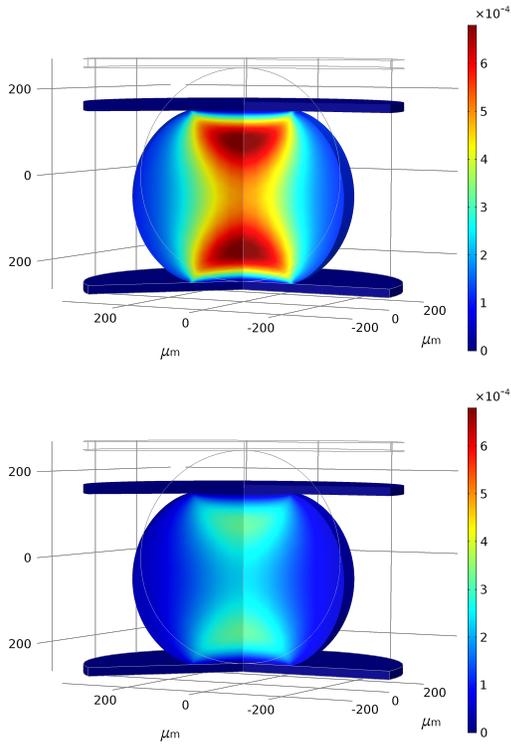

**Figure 8**. The upper figure shows the von Mises stress (in $\mu N/\mu m^2$) directly after the initial compression of the aggregate, i.e. at $t = 0.2s$. The lower figure shows how the von Mises stress has been released at $t = 1497s$ due to the viscoelasticity.

Note that we were able to simulate much larger displacements (up to 30% of the original height) in a static regime. Nevertheless, it has proven difficult to find suitable settings for the numerical solver in the transient regime, particularly concerning the penalty for the contact solver.

Having this model at hand, it is now easy to reproduce the compression results from [4].
In one of the experiments shown there, the upper plate was displaced by $50\mu m$. Then, the stress is measured via the formula
$$F_c = (p + \sigma)A_{mid}$$
where $F_c$ is the contact force, $p$ the hydrostatic pressure, $\sigma$ the elastic stress and $A_{mid}$ the cross-section area of the spheroid's midplane. For simplicity, we will combine the two stresses and just measure $F_c/A_{mid} = \tilde{\sigma}$, i.e. we do not account for any residual stresses. In our numerical realization, the displacement is performed within the first $0.2s$ of the simulation. The contact force $F_c$ is calcululated by integrating the contact pressure $p_c$ with respect to the contact area, which we determine by the criterion $\{p_c > 10^{-8}\}$.

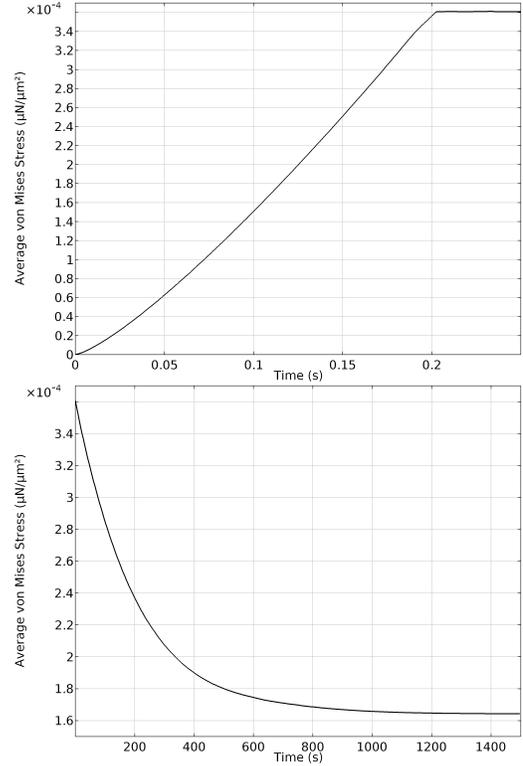

**Figure 9**. Average von Mises stress versus time. The upper figure shows how the stress builds up within the first $0.2s$. The lower figure shows how the stress is released over time.

For the parameter choice $C = 150Pa, \alpha = 0.2, \beta_1 = 0.4, \tau_1 = 200s$, we are able to qualitatively reproduce the stress relaxation curve from [4, Fig. 3], see Figure 10. The black line denotes the approximate stress in $\mu N/\mu m^2$, whereas the blue circles represent the experimental data from [4, Fig. 3]. We have extracted these data using WebPlotDigitizer[2].

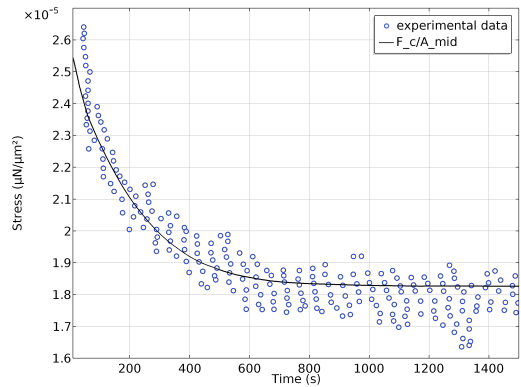

**Figure 10**. Numerically approximated stress versus time (black line) and experimental data from [4] (blue circles).

[2]http://arohatgi.info/WebPlotDigitizer/app/

If one was interested in actually matching the experimentally obtained data, it would be straightforward to employ the "Optimization" module to determine the optimal values for the four parameters $C, \alpha, \beta_1, \tau_1$ in order to match the measurements, see e.g. [12].

**Conclusions**

In this article, we have given two examples how the COMSOL Multiphysics "Structural Mechanics" module can be used to perform biomechanical tissue simulations. Based on the established approach, it is now straightforward to apply COMSOL to other biomechanical biomechanical tissue simulations. In particular, COMSOL might be used to obtain parameter values from experimentally measured data and, based on the estimated parameter values, to perform predictions with the aid of numerical simulations.

## Acknowledgements

We thank our colleagues for discussion and the COMSOL support staff for their excellent support, especially Jonas Lejon, Thierry Luthy and Zoran Vidakovic. This work has been supported by the Swiss National Science Foundation through the SNF Sinergia grant 170930.